\documentclass[aps,twocolumn,superscript]{revtex4}

\usepackage{graphicx}

\usepackage[dvipsnames]{xcolor}
\usepackage{amsmath}
\usepackage{hyperref}
\usepackage{gensymb}
\hypersetup{colorlinks=true,
            linkcolor=blue,
            citecolor=blue,
            urlcolor=orange}
\usepackage[resetlabels,labeled]{multibib}            
\newcites{App}{App Readings}

\begin{document}

\title{Splitting of electronic spectrum in paramagnetic phase of 
itinerant \\ferromagnets and altermagnets}
\author{A. A. Katanin}
\affiliation{Center for Photonics and 2D Materials, Moscow Institute of Physics and Technology, Institutsky lane 9, Dolgoprudny, 141700, Moscow region, Russia}
\affiliation{M. N. Mikheev Institute of Metal Physics of the Ural Branch of the Russian Academy of Sciences, S. Kovalevskaya Street 18, 620990 Yekaterinburg, Russia}

\begin{abstract}
We study self-energy effects induced by strong magnetic fluctuations in the paramagnetic phase of strongly-correlated itinerant magnets within the density functional theory combined with the dynamical mean field theory (DFT+DMFT approach) and its non-local extension. We show that both local and non-local magnetic correlations yield a splitting of the electronic spectrum in the paramagnetic phase, such that it closely resembles the DFT band structure in the ordered phase. We demonstrate these effects on $\alpha$-iron, half-metal CrO$_2$, van der Waals material CrTe$_2$, and altermagnet CrSb. Although the obtained split bands do not possess a certain spin projection, 
their splitting suppresses spectral weight at the Fermi level. Even when originating from local magnetic correlations, the splitting is strongly momentum dependent as a consequence of the orbital selectivity of non-quasiparticle states.  The relative importance of non-local vs. local correlations depends on the proximity to half filling of $d$ states: closer to half filling, the role of local correlations increases. 
\end{abstract}

\maketitle


The first explanation of itinerant magnetism was given by E. Stoner in 1938 \cite{Stoner} and it was suggested that ferromagnetism in metals is accompanied by the splitting of electronic bands due to the exchange interaction's contribution to the free energy, proportional to the square of magnetization. This contribution was shown to originate from the on-site Coulomb repulsion \cite{Hubbard}, as it is also described in many textbooks \cite{Moriya,vonsovsky,mattis,yosida,white}. Although in strongly correlated systems correlation effects may in general lead to the renormalization of bands, in the ferromagnetically ordered state these effects are suppressed by the band splitting; therefore, the Stoner picture remains qualitatively applicable.

The situation is different in the paramagnetic phase even sufficiently close to the magnetic transition. While in two-dimensional systems the so-called quasi-splitting of Fermi surfaces may occur \cite{QS1,QS1a,QS2,QS3,QS4}, in three-dimensional systems the magnetic correlations are generally weaker and not sufficient to establish a pronounced change in the electronic spectrum. At the same time, two effects appear to be important in the paramagnetic phase. The first effect is the presence of local correlations, which originate mainly from Hund interaction in the so-called Hund metals \cite{Hund1,Hund2,Hund3,Hund4}. They lead to the formation of local magnetic moments and non-Fermi-liquid behavior and/or a non-quasiparticle form of some of the electronic bands. Notably, many “classical” magnets, e.g. iron \cite{OurFe,Sangiovanni}, as well as many other transition metals and their compounds, belong to the class of Hund metals. Although the importance of this effect for quasi-splitting of the Fermi surfaces was discussed previously for the two-orbital model \cite{QS2}, the study of its effect in multi-orbital is of certain interest.

The second effect is the presence in many magnetic systems of the so-called extended van Hove singularities, which represent lines of van Hove singularities and yield strong peaks in the density of states \cite{KatsVH}. These singularities effectively reduce the dimension of the system and may lead to a stronger renormalization of the electronic spectrum by the non-local (e.g. magnetic) correlations. In our opinion, the effect of these singularities on the renormalization of the electronic spectrum in the paramagnetic phase has not been investigated in detail. Experimentally, electronic spectrum splitting above the Curie temperature was observed in thin films of iron and SrRuO$_3$ \cite{FeSpl,SrRuO3,SrRuO3-1,SrRuO3-2}, in cobalt \cite{Co}, as well as in single-layer CrTe$_2$ \cite{CrTe2exp}.

In the present paper, we consider the effects of local and non-local contributions to self-energy in several metallic magnets. We consider magnets belonging to  different classes, but all having extended van Hove singularities close to the Fermi level. In particular, we investigate the effects of the self-energy in strong magnet ($\alpha$-iron), half metal (CrO$_2$), van der Waals magnet CrTe$_2$, and altermagnet CrSb. We show that the effects of the local self-energy are pronounced only very near half filling (which is the case of CrTe$_2$ and CrSb), while the non-local self-energy effects are pronounced for systems further away from half filling (i.e., for iron and CrO$_2$). In combination, in all considered systems, these two effects yield splitting of the electronic spectrum near the van Hove singularity, which is strikingly qualitatively similar to that obtained within the DFT calculation in the ferromagnetic phase, although the obtained bands do not possess a certain spin orientation.

To describe the effect of local correlations, we use the standard DFT+DMFT approach \cite{DMFT_rev,anisimov1997,LK1998,DFTplusDMFT}. The non-local self-energy effects  in the multi-orbital systems can be described within the ab initio dynamic vertex approximation (D$\Gamma$A) \cite{AbInitioDGA0,AbinitioDGA,AbinitioDGA1}, which is an extension of the previously proposed single-band D$\Gamma$A \cite{DGA,DGA1,DGA2}. The multi-orbital approach, however, was previously derived for SU(2) symmetry of Hund interaction, the consideration of which is rather computationally time-consuming. We therefore consider its formulation and implementation for Ising symmetry of Hund interaction, which is obtained with the help of the recently proposed unambiguous fluctuation decomposition of the self-energy \cite{Krien}.



{\it General formalism}. We consider the DFT+DMFT approach with the tight-binding multi-orbital Hubbard model 
\begin{equation}
{H} = \sum_{{\bf k},\lambda\lambda',\sigma}H_{\bf k}^{\lambda\lambda'} c^+_{{\bf k}\lambda\sigma} c_{{\bf k}\lambda'\sigma}^{} +H_{\rm int}  - {H}_{\rm DC},
\end{equation}
describing electrons that are moving on the sites of the lattice ($c^+_{\mathbf{k}\lambda\sigma}$ and $c_{\mathbf{k}\lambda\sigma}$ are the electron creation and destruction operators, $\lambda,\lambda'$ are the orbital indices), subject to the on-site Coulomb repulsion $H_{\rm int}$, and $H_{\rm DC}$ is the double counting contribution, which is needed to keep the ab initio quasiparticle energies, obtained by diagonalizing $H_{\bf k}^{\lambda\lambda'}$ unchanged by the interaction. For the Coulomb repulsion, we consider a density-density Hamiltonian
\begin{equation}
H_{\rm int}=\frac{1}{2}\sum\limits_{i,mm',\sigma\sigma'} U^{mm^\prime}_{\sigma\sigma^\prime}
{n}_{im\sigma} {n}_{im^\prime\sigma^\prime},
\label{Hint}
\end{equation}  
where $n_{im\sigma}=c^+_{im\sigma} c_{im\sigma}$ (the indices $m$,$m'$ numerate $d$-orbitals). 
The local self-energy $\Sigma^{{\rm loc},m}_{i\nu}$, obtained within DMFT, is used to obtain the DMFT non-local Green's function $G_k^{mm'}=[i\nu_n+\mu+M_{\rm DC}-H_{\mathbf k}-\Sigma^{\rm loc}_{i\nu}]^{-1}$ where $M_{\rm DC}$ is the double counting contribution to the self-energy, and $k=( {\mathbf k},i\nu_n)$ is the wave vector and frequency multi-index. 

The non-local self-energy is obtained as (see Supplemental Material \cite{SM})
\begin{align}
\Sigma&_k^{mm'} =\Sigma _{i\nu }^{\mathrm{loc,}mm^{\prime }}+\frac{1}{2}T\mathcal{N}_{l}\sum\limits_{q,\widetilde{m}}\left[ \gamma
_{c,\nu q}^{m\widetilde{m}}W_{c,q}^{\widetilde{m}m^{\prime
}}\right.\notag\\
&\left.+3\gamma
_{s,\nu q}^{m\widetilde{m}}W_{s,q}^{\widetilde{m}m^{\prime
}}\right] G_{k+q}^{mm'}\label{SEnonloc}
\end{align}
where the first and second terms in the square bracket describe the contribution of charge and spin fluctuations, $\chi _{k^{\prime }q}^{0,m^{\prime }\widetilde{m}%
^{\prime}}=-TG_{k'}^{m'\tilde{m}'}G_{k'+q}^{\tilde{m}'m'}$ is the bare bubble, $W_{c(s)}^{mm'}$ is the renormalized charge (spin) interaction, and $\gamma_{c(s),\nu q}^{m\widetilde{m}}$ are the triangular Hedin vertices (see Ref. \cite{SM}),  $\mathcal{N}_l..$ stands for the non-local part of the following expression. 

For the vectors ${\mathbf k}$ located near the Fermi surface, the Green's  function $G_{k+q}$ is enhanced at the flat parts of the electronic dispersion and, as we discuss in detail in the Supplemental Material\cite{SM}, one of its eigenvalues fulfills on the real frequency axis $G_{(\mathbf{k}_F+{\bf q},\nu)}\propto 1/(\nu-i\mathrm{Im}\Sigma^{\rm loc}_\nu)$. Since its dependence on ${\mathbf q}$ is weak in this case, this behavior is translated to the non-local self-energy, which, as we also demonstrate below by explicit numerical calculations, has a {\it positive} derivative $\partial {\rm Re}\Sigma_{({\mathbf k}_F,\nu)}/\partial \nu$, breaking the quasiparticle concept, similarly to the previous discussion of the single-band case \cite{QS1,QS1a,QS4}. 
In this case, the non-local self-energy effects in the static approximation are therefore controlled by the 
averaged interaction $\sum_{\mathbf q} W_{{\mathbf q},i\omega_n=0}$. This behavior competes, however, with the local self-energy effects, which are determined by the respective local spin interaction $W_{\rm loc,\omega=0}$.

{\it Numerical results}. 
For DFT calculations, we use the Quantum Espresso package \cite{QE} with ultrasoft pseudopotentials from the SSSP PBEsol Precision library \cite{PPP}, and subsequent Wannier projection using maximally localized Wannier functions obtained within the Wannier90 package \cite{Wannier90}. The DFT+DMFT approach to study exchange interactions in the three former metallic ferromagnets was considered in Refs.  \cite{OurJq,MyCrO2,CrTe2}. We use a parametrization of the Coulomb interaction by Slater integrals $F^i$ with around-mean-field form of the double counting correction \cite{AMF}. 
For DMFT calculations, we use the segment continuous time Monte-Carlo (CT-QMC) \cite{CT-QMC} \textsc{iQIST} solver \cite{iQIST}, which allows us to considerably reduce the stochastic noise and obtain details of the frequency dependence of the vertex functions over a sufficiently broad frequency range. We also use corrections for finite frequency boxes in the Bethe-Salpeter equations \cite{My_BS}; further details can be found in Refs. \cite{OurJq,MyCrO2,CrTe2}. The obtained Green's functions, containing local or non-local self-energies, are analytically continued to the real axis using \textsc{ana\_cont} package \cite{AnaCont}.

\begin{figure}[b]
\centering
\includegraphics[width=1.0\linewidth]{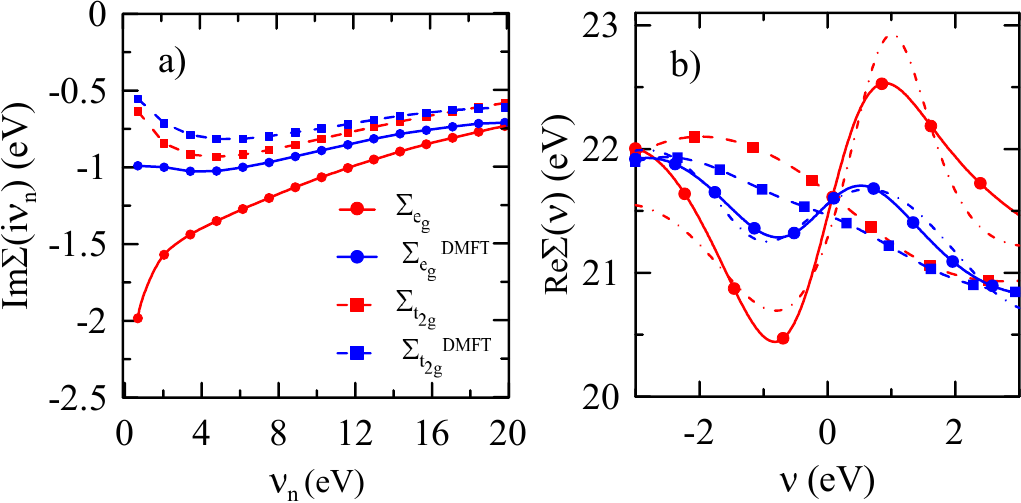}
\caption{(Color online) Frequency dependence of the imaginary part of electronic self-energy of iron at the imaginary frequency axis (a) and real part of the self energy at small energies of the real frequency axis (b) in DFT+DMFT approach (blue symbols) and the approach considering non-local correlations (red symbols) at the $P$ point of the Brillouin zone at $\beta=4.6$~eV$^{-1}$. Circles (squares) mark self-energies of $e_g$ ($t_{2g}$) orbital states. Solid and dashed lines in (b) show the result of analytical continuation with \textsc{ana\_cont}, dot-dashed lines correspond to
Pade approximants for $e_g$ states.}  
\label{FigFeSgm}
\end{figure}

\begin{figure*}[t]
\centering
\includegraphics[width=0.8\linewidth]{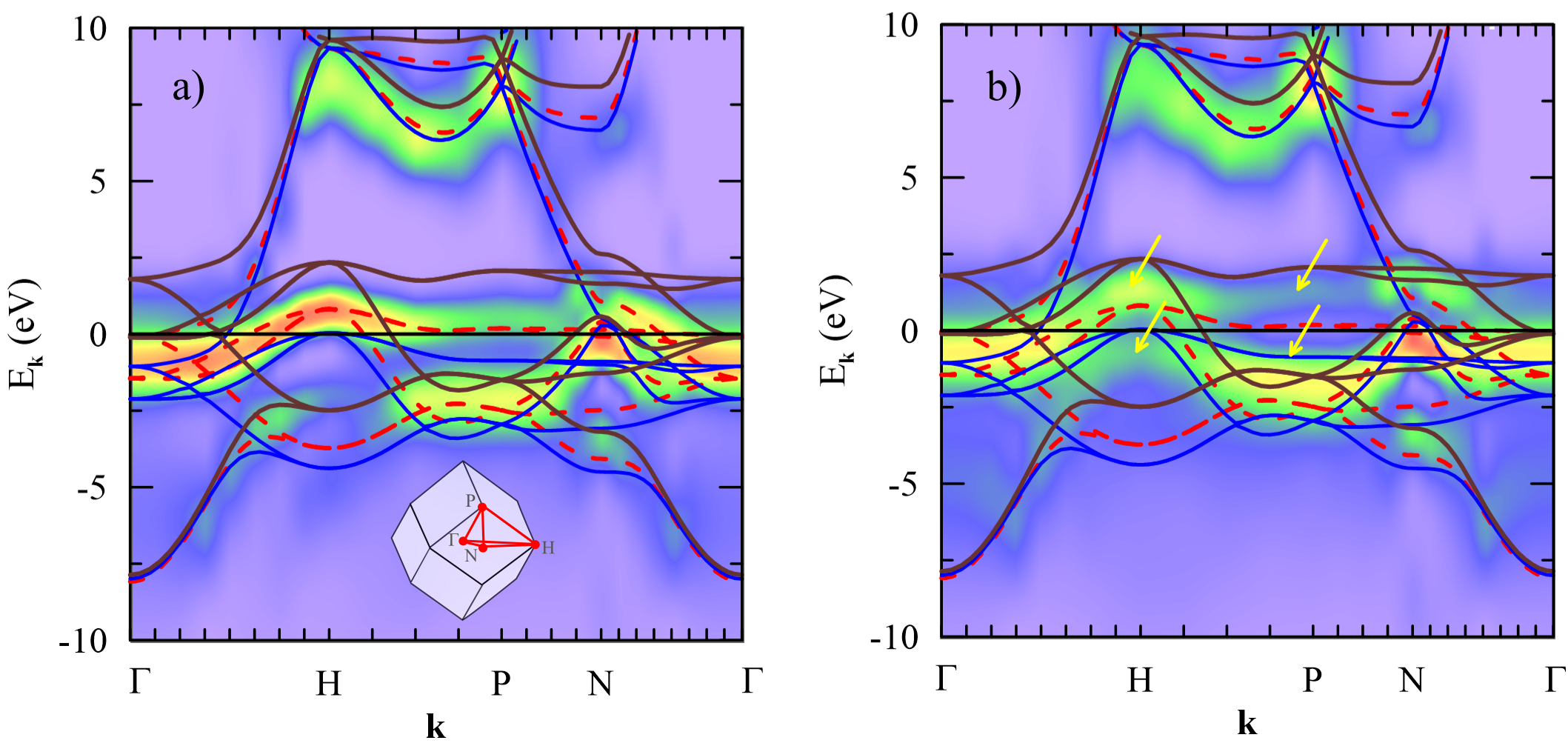}
\caption{(Color online) Band structure of iron in the ferromagnetic (brown and blue solid lines for different spin projections) and non-magnetic (red dashed lines) phases, compared to the spectral density in DMFT approach (a) and the approach, accounting for the non-local corrections to the self-energy (b) at $\beta=4.6$ eV$^{-1}$. Yellow arrows in (b) show position of bands, formed by spin fluctuations in paramagnetic phase. Inset in (a) shows the path in the Brillouin zone.}  
\label{FigFe}
\end{figure*}

We consider first bcc iron as a strongly-correlated itinerant magnet. In the Wannier projection, we consider $3d$, as well as $4s,4p$ states. The DMFT approach is applied at $U=F^0=4$~eV and Hund exchange $J=(F^2+F^4)/14=0.9$~eV, considered previously in Refs. \cite{OurJq,MySU2}. We consider first the results at $\beta=1/(k_B T)=4.6$~eV$^{-1}$ ($T=2900$~K). This relatively high temperature is chosen in view of high Curie temperature of iron in the density-density approach, which corresponds to $\beta_c=4.87$~eV$^{-1}$($T_C=2380$~K). The respective uniform susceptibility at the considered temperature $\chi_{{\mathbf q}=0}=105$~eV$^{-1}$ is much larger than the local transverse susceptibility  $\chi_{\rm loc}=12$~eV$^{-1}$. In Fig. \ref{FigFeSgm} we show the comparison of the DFT+DMFT electronic self-energy $\Sigma _{\nu }^{\mathrm{loc,}m}$ (which is momentum-independent) and the non-local electronic self-energy obtained from Eq. (\ref{SEnonloc}) at the $P$ point of the Brillouin zone, where the effect of the extended van Hove singularity is the strongest. One can see that the non-local correlations strongly enhance the non-quasiparticle behavior of $e_g$ states, which mainly contribute to the extended van Hove singularity, as discussed previously within the DFT+DMFT approach in Ref. \cite{OurFe}. In particular, we find positive derivative $\partial {\rm Im}\Sigma^{e_g}({\mathbf k}_P,i\nu)/\partial \nu>0$, which translates into $\partial {\rm Re}\Sigma^{e_g}({\mathbf k}_P,\nu)/\partial \nu>0$ at the real frequency axis (see the result of analytic continuation in Fig. \ref{FigFeSgm}b), in contrast to the $t_{2g}$ states, having opposite sign of derivatives. The magnitude of the derivative is strongly enhanced by non-local correlations.

\begin{figure}[b]
\centering
\includegraphics[width=0.95\linewidth]{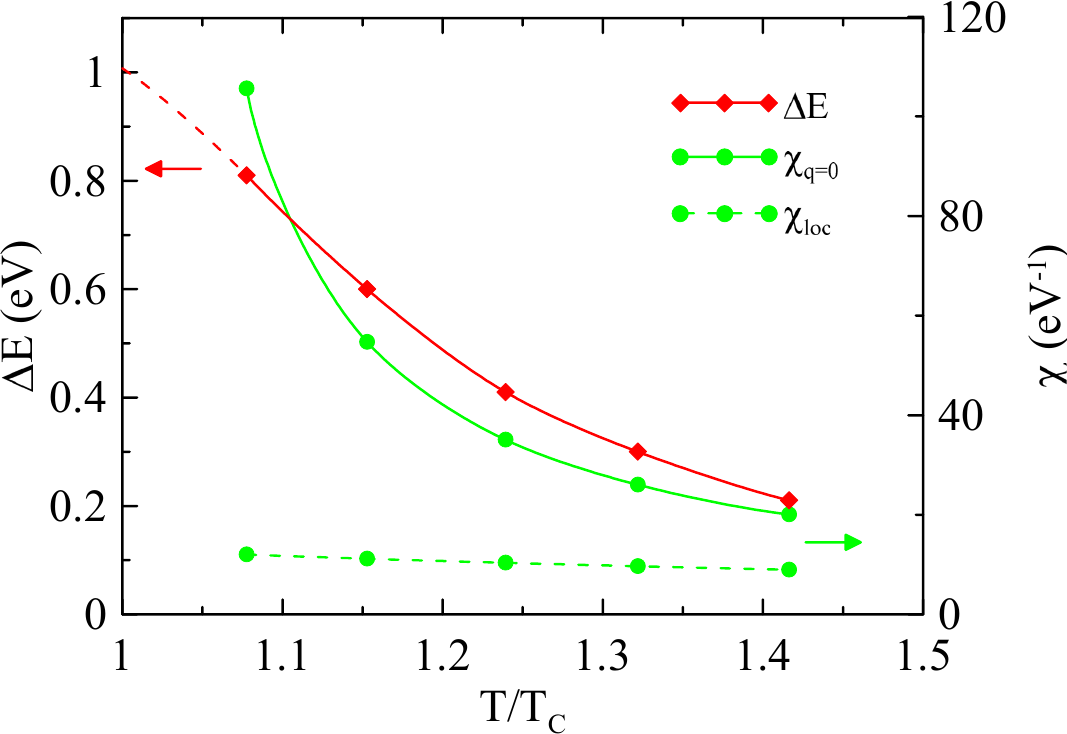}
\caption{(Color online) Temperature dependence of the electron spectrum splitting in iron at the point $P$ of the Brillouin zone (red lines with rhombs, left axis) compared to the temperature dependence of uniform (green solid line with circles) and local (green dashed line with circles) susceptibilities (right axis). Dashed line shows the result of the extrapolation of splitting to Curie temperature.}  
\label{FigFeSpl}
\end{figure}

In Fig. \ref{FigFe} we show the effect of this non-quasiparticle behavior on the spectral density in DFT+DMFT approach and the non-local self-energy result, in comparison to the DFT band structure of iron in non-magnetic and ferromagnetic phases.  One can see (Fig. \ref{FigFe}a) that the spectral density in DMFT approach is sufficiently close to the DFT band structure in the non-magnetic phase, with a slight shift of the flat band in the direction $H$-$P$-$N$, induced mainly by a small $\mathrm {Re}\Sigma^{\rm loc}_0-M_\mathrm{DC}$ correction, and to a lesser degree, the above discussed weakly non-quasiparticle behavior of $e_g$ states.  
The non-local effects (see Fig. \ref{FigFe}b), yielding strongly non-quasiparticle behavior of $e_g$ states, result in a qualitative change of the band  structure, which becomes remarkably similar to the DFT band structure in the ferromagnetic phase. 
In particular, we observe splitting of the band structure along the $H$-$P$-$N$ direction in momentum space (where the extended van Hove singularity is present).  This splitting is reminiscent of the earlier discussed quasi-splitting in the single-band two-dimensional systems \cite{QS1,QS1a,QS2,QS3,QS4}, but it differs from that case by both dimensionality and multi-band electronic structure. 

The obtained temperature dependence of the splitting at the $P$ point (estimated as a difference $\Delta E$ between the position of the band above the Fermi level and its value in the DFT+DMFT approach) is shown in Fig. \ref{FigFeSpl}. One can see that the obtained splitting of electronic structure persists in a rather wide temperature range. 
At low temperatures, the splitting (determined by $\sum_{\mathbf q} W_{s,({\mathbf q},0)}$) increases slower than the uniform susceptibility. In particular, the splitting reaches a finite value $\Delta E\simeq 1$~eV at $T\simeq T_C$ and is expected to approach smoothly the low-temperature value $(\Delta E)_{\rm sat}\simeq 2$~eV below the Curie temperature.  We note that the exchange splitting in thin iron films was observed above the Curie temperature by measuring the splitting of image-potential surface states using spin-resolved two-photon photoemission in Ref. \cite{FeSpl}. Although the magnitude of the splitting of image-potential surface states is expected to be much smaller than the exchange splitting of electronic states, the obtained temperature dependence shows a qualitative similarity to the experimental splitting, decreasing, however, weaker with temperature. This weaker decrease can be related to the overestimate of the splitting due to the approximations made (e.g., using local self-energy in the internal Green's functions, density-density approximation, etc.).

As we discuss in Supplemental Material \cite{SM}, similar to iron splitting of the electronic spectrum is expected for the half metallic compound CrO$_2$. In this case, DMFT yields a somewhat stronger change in the electronic band structure than in iron, which makes the band structure similar to that obtained within DFT for the ordered phase. Considering the effect of the non-local self-energy yields a stronger splitting of bands, resulting again in qualitative agreement with the ferromagnetic DFT band structure. The stronger effect of the local self-energy in CrO$_2$ in comparison to iron can be attributed to the somewhat closer proximity to half-filling of $d$ states ($n_d^{{\rm CrO}_2}=3.76$ vs. $n_d^{\rm Fe}=6.4$).

\begin{figure}[t]
\centering
\includegraphics[width=1.00\linewidth]{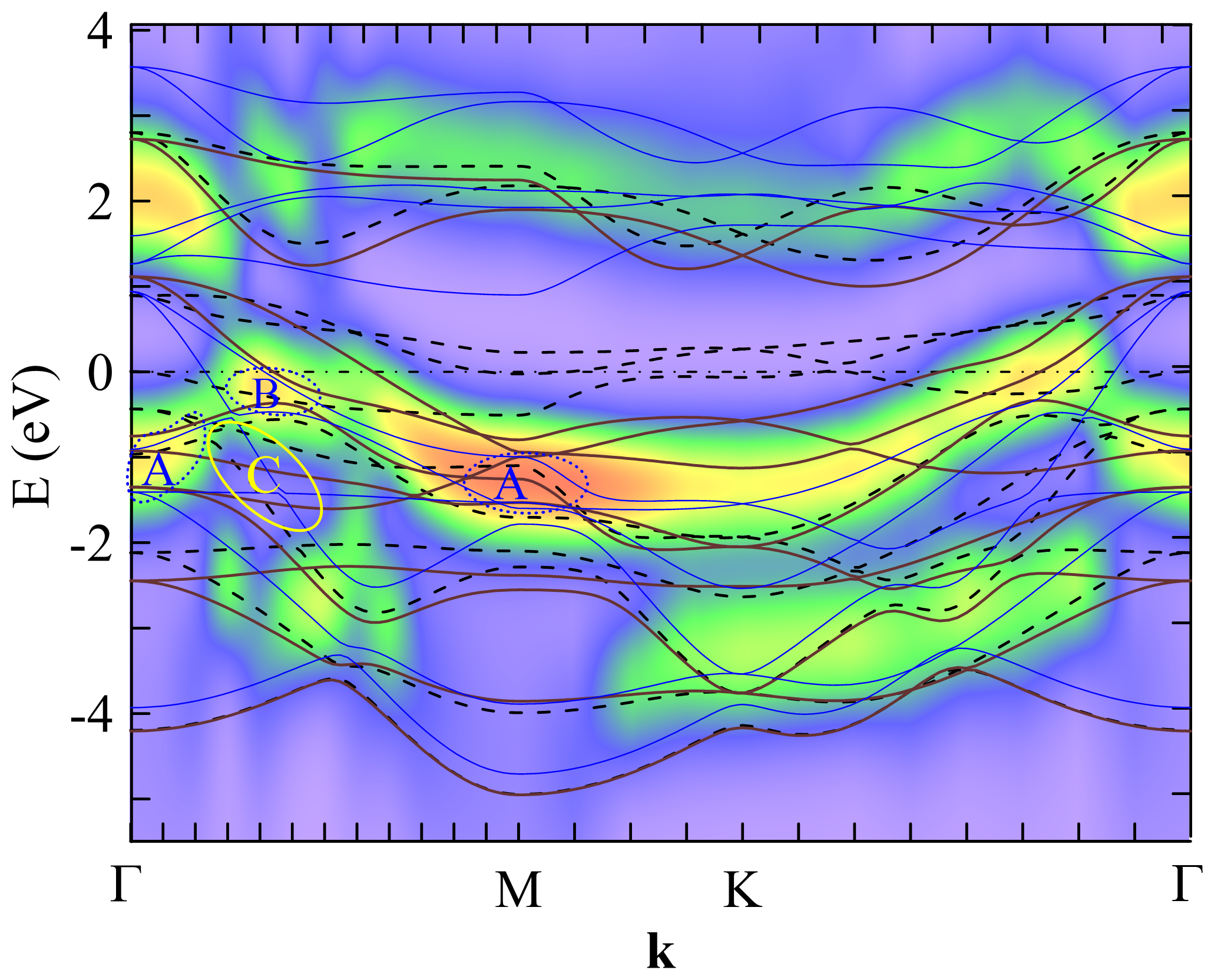}
\caption{(Color online) Band structure of monolayer CrTe$_2$ in the ferromagnetic (violet and blue solid lines for different spin projections) and non-magnetic (black dashed lines) phases, compared to the spectral density in the approach, which account for the non-local correlations at T=1160~K ($\beta=10$~eV$^{-1}$). The regions $A$,$B$ bounded by dotted lines may be related to the peaks in ARPES with $s$,$p$ polarized photons in Ref. \cite{CrTe2exp}, and originate from local magnetic correlations. The spectral weight is suppressed by non-local correlations in the area C.}    
\label{Fig:CrTe2_sl}
\end{figure}

In Fig. \ref{Fig:CrTe2_sl} we present the results for the monolayer van der Waals magnet CrTe$_2$. We choose the lattice parameter $a=3.81$~\AA~and in the Wannier projection we use Cr $3d$, as well as Te $5p$ states. We consider $T=1160$~K ($\beta=10$~eV$^{-1}$) which is well above the DMFT magnetic transition temperature $T_c\simeq500$~K \cite{CrTe2_Erratum} ($\beta_c=23.2$~eV$^{-1}$; we choose $U=F^0=2.8$~eV, $J=0.9$~eV, corresponding to $U_K=3.9$~eV, $J_K=0.65$~eV in Kanamori parametrization \cite{cRPA,CrTe2}). We obtain the uniform susceptibility $\chi_{{\bf  q}=0}=92$~eV$^{-1}$ and $\chi_{\rm loc}=57$~eV$^{-1}$.  Although for monolayer CrTe$_2$ the quasi-splitting of the Fermi surface originating from magnetic correlations  proposed previously for two-dimensional materials \cite{QS1,QS1a,QS2,QS3,QS4} could be active, 
we find strong band splitting already in DFT+DMFT approach (see Supplemental Material \cite{SM}), which therefore originates from the local magnetic correlations. This can be attributed to the relatively large local susceptibility (the respective maximal in orbital indexes local interaction $\gamma_{s,\nu_1,{\rm loc}}W_{s,{\rm loc}}\simeq 4.5$~eV at the first Matsubara frequency $\nu_1$ is comparable to the averaged {\it non-local} interaction in iron, $\sum_{\mathbf q}{\gamma_{s,\nu_1,({\mathbf q},0)}W_{s,({\mathbf q},0)}\simeq 4.3}$~eV, and substantially larger than the {\it local} interaction in iron, $\gamma_{s,\nu_1,{\rm loc}}W_{s,{\rm loc}}\simeq 1.3$~eV at approximately the same $\chi_{{\mathbf q}=0}$), which is related to the proximity to half filling of $d$ states in CrTe$_2$ ($n_d=4.85$). In contrast to the metal-insulator transition at integer filling, we find this splitting strongly momentum dependent, despite the momentum independence of the local self-energy, which is due to contribution of only certain orbital states to the flat part of the electronic spectrum.
In view of the weaker dependence of local magnetic susceptibility, compared to the non-local one, on temperature, we expect that band splitting in this case can be observed in a broader temperature range. The effect of the non-local self-energy is sufficiently small in this case, suppressing however spectral weight at some points of the Brillouin zone.

Recently, the band splitting above the Curie temperature for single-layer CrTe$_2$, deposited on bilayer graphene,  was observed by angle-resolved photoemission (ARPES) in Ref. \cite{CrTe2exp}. In Fig. \ref{Fig:CrTe2_sl} we mark by $A$ and $B$ the maxima of the spectral functions, which seem to be related to the maxima of the ARPES, originating from the $s$ and $p$ polarized photons, observed in Ref. \cite{CrTe2exp}. Although we do not find the spectrum splitting of the peaks corresponding to the $p$ polarization of photons (likely because of the high considered temperature), we emphasize the suppression of the spectral weight by non-local correlations in the region $C$, which is near the regions $A$ and $B$.  
  As we discuss in Supplemental Material \cite{SM}, the results for bulk CrTe$_2$ are similar to those for the monolayer compound.  

\begin{figure}[t]
\centering
\includegraphics[width=0.95\linewidth]{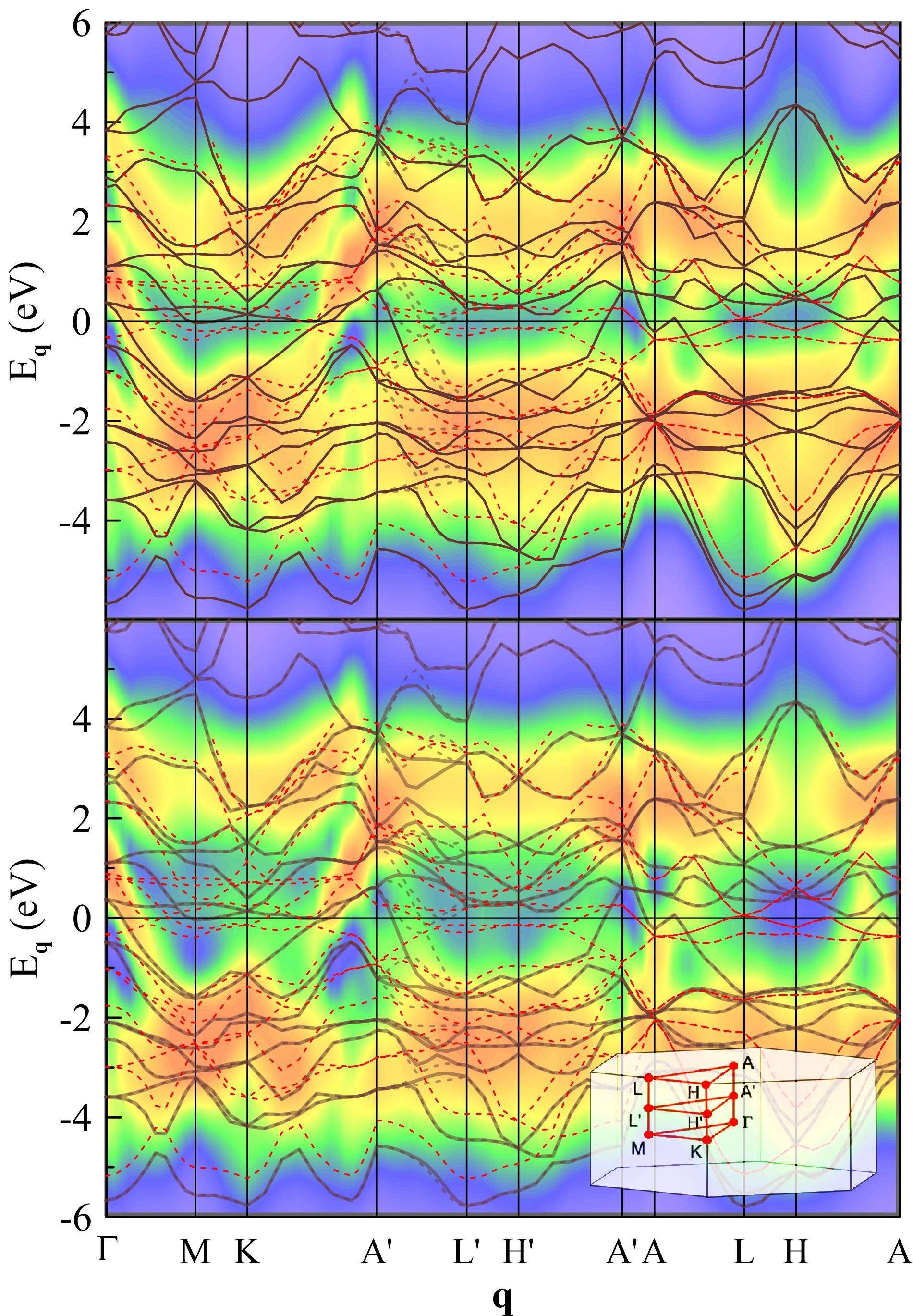}
\caption{(Color online) Band structure of altermagnet CrSb in the magnetic  (dark solid and dashed lines for different spin projections) and non-magnetic (red dashed lines)  phases, compared to the spectral density in DMFT approach (top) and the approach, accounting the non-local corrections to the self-energy (bottom) at $\beta=5$ eV$^{-1}$.}   
\label{FigCrSb}
\end{figure}

Finally, we consider the electron spectral density of an altermagnet CrSb. This compound has two Cr sites per unit cell, which tend to order antiferromagnetically. Yet, these two sublattices are connected by the operation $\tilde{C}_6=C_6 {\mathbf t}$, combining the six-fold rotation $C_6$ with the translation ${\mathbf t}=(0,0,c/2)$ where $c$ is the lattice parameter along $z$ axis. In the symmetry broken state, this yields spin splitting at $k_z\neq 0,\pi/c$ \cite{CrSb1,CrSb2,CrSb3,CrSb4}. To visualize band structure, we choose the path, containing the points $A'=(0,0,4\pi/(9c))$ and $L'=2\pi(0,1/(\sqrt{3}a),2/(9c))$. We use the same Coulomb interaction strength as for CrTe$_2$ and present the spectral functions at $\beta=5$~eV$^{-1}$ in Fig. \ref{FigCrSb}, comparing them to DFT calculations in antiferromagnetic and nonmagnetic phases. The temperature in DFT+DMFT calculation is also chosen close to the Neel temperature $\beta_N=5.9$~eV$^{-1}$ ($T_N=1950$~K); the respective susceptibilities at the considered temperature $\chi_{\mathbf q=0}^{11}=95$~eV$^{-1}$ and $\chi_{\mathbf q=0}^{12}=-64$~eV$^{-1}$ (the upper indices numerate Cr sites within the unit cell, the negative value of $\chi_{\mathbf q}^{12}$ indicates antiferromagnetic correlations between Cr sites belonging to the same unit cell), whose absolute values are larger than the local susceptibility $\chi_{\rm loc}=34$~eV$^{-1}$. 

One can see that both the DFT+DMFT approach and the approach considering the non-local corrections to the self-energy lead to the splitting of the electronic spectrum of CrSb. Similarly to CrTe$_2$, we attribute the large effect of local correlations to the large local susceptibility, which originates from proximity to half filling ($n_d=4.9$). As for CrTe$_2$, DFT+DMFT approach yields results quite similar to those in the ordered phase of the DFT approach. This is also pronounced in the $A'$-$L'$ direction, where the spin splitting of the electronic spectrum is present in the DFT approach in the antiferromagnetic phase. The approach considering non-local self-energy corrections yields even larger band splitting.


In conclusion, we have analyzed the effect of the local and non-local self-energy effects on the electronic spectrum of strongly correlated itinerant magnets in the paramagnetic phase. We have obtained sufficiently strong band splitting near van Hove singularities of the electronic spectrum. For the magnets with a filling of $d$ states not very close to half filling (e. g., iron and CrO$_2$), we find pronounced effects of non-local magnetic correlations in the vicinity of the Curie temperature. At the same time, for magnets with the $d$ states filling closer to half filling (e.g. CrTe$_2$ and CrSb) we find strong splitting of the electronic spectrum already due to local spin correlations. In contrast to the metal-insulator transition, this splitting is strongly momentum dependent, which is due to the contribution of certain orbital states to the flat part of the electronic spectrum. Since the local correlations decay slower with the increase of temperature, this effect is expected to be present in a much broader temperature range. We emphasize the qualitative agreement of the obtained spectral features with those observed above the Curie temperature in thin films of iron \cite{FeSpl}, as well as in single-layer CrTe$_2$ \cite{CrTe2exp}. 

The obtained split energy bands do not possess any certain spin polarization, although they each carry only part of the spectral weight of the non-magnetic bands, which brings their similarity to Hubbard subbands. However, in contrast to Hubbard subbands, they are dispersive; therefore, they combine itinerant and local moment features. These bands are expected to also affect transport properties (which is, however, the subject of a separate study) due to the suppression of the spectral weight at the Fermi level. We also expect that these states are highly sensitive to magnetic field. In particular, due to high magnetic susceptibility, even small magnetic fields yield strong magnetic polarization, which corresponds to having a preferable spin projection of the obtained bands. This may allow control over the spin polarization of electronic states by a weak magnetic field, without strong changes in the electronic band structure, which also represents an interesting subject of future studies and may find its application in magnetic nano-devices and spintronics.

The considered density-density approximation overestimates the magnetic transition temperatures \cite{Sangiovanni,MySU2,AnisimovSU2}; it is also expected to somewhat overestimate the renormalization of the electronic spectrum \cite{QS3}. For further studies, the extension of the consideration of the present paper to the SU(2) symmetric Hund interaction is of certain interest, since it may allow one to consider magnetic transition temperatures comparable to those for experimental compounds. 
Also, further study of the effect of the non-local self-energy  and vertex corrections on the obtained splitting and its impact on the other physical observables is of certain interest.

\section*{Acknowledgements} 
The author is grateful to I. Dedov and I. Goremykin for discussions. Performing the DMFT calculations was supported by the Russian Science Foundation
(project 24-12-00186). 
DFT calculations were carried out within the framework of the state assignment of the Ministry of Science and Higher Education of the Russian Federation for the IMP UB RAS. The calculations were
performed on the cluster of the Laboratory of Material
Computer Design of MIPT.

\clearpage
\appendix
\widetext

\renewcommand\theequation{S\arabic{equation}}
\renewcommand\thesubsection{\arabic{subsection}}
\renewcommand\thefigure{S\arabic{figure}}
\setcounter{equation}{0}
\setcounter{figure}{0}
\setcounter{page}{1}

\section*{SUPPLEMENTAL MATERIAL \\for the paper ``Splitting of electronic spectrum in paramagnetic phase of 
itinerant \\ferromagnets and altermagnets"}

\subsection{Derivation of Eq. (\ref{SEnonloc})}

We consider the local interaction part of the Hamiltonian in the most general
form%
\begin{equation}
H_{\mathrm{int}}=-\frac{1}{4}\sum\limits_{\alpha \beta \gamma \delta
}\sum\limits_{i,mm^{\prime }m^{\prime \prime }m^{\prime \prime \prime
}}U_{\alpha \beta \gamma \delta }^{mm^{\prime }m^{\prime \prime }m^{\prime
\prime \prime }}c_{im\alpha }^{+}c_{im^{\prime }\beta }^{+}c_{im^{\prime
\prime }\gamma }^{{}}c_{im^{\prime \prime \prime }\delta }^{{}},
\label{Hint1}
\end{equation}%
where $\alpha ,\beta ,\gamma ,\delta =\uparrow ,\downarrow $ are spin
indices, $m,m^{\prime },m^{\prime \prime },m^{\prime \prime \prime }$ are
indices of $d$-orbitals. The conservation of the $z$-component of electron
spin projection together with antisymmetry $U_{\alpha \beta \gamma \delta
}^{mm^{\prime }m^{\prime \prime }m^{\prime \prime \prime }}=-U_{\alpha \beta
\delta \gamma }^{mm^{\prime }m^{\prime \prime \prime }m^{\prime \prime }}$
requires 
\begin{equation}
U_{\alpha \beta \gamma \delta }^{mm^{\prime }m^{\prime \prime }m^{\prime
\prime \prime }}=U_{\alpha \beta }^{mm^{\prime }m^{\prime \prime }m^{\prime
\prime \prime }}\delta _{\alpha \gamma }\delta _{\beta \delta }-U_{\alpha
\beta }^{mm^{\prime }m^{\prime \prime \prime }m^{\prime \prime }}\delta
_{\alpha \delta }\delta _{\beta \gamma }.
\end{equation}%
The Hamiltonian (\ref{Hint1}) can be then reduced to 
\begin{equation}
H_{\mathrm{int}}=\frac{1}{2}\sum\limits_{\alpha \beta
}\sum\limits_{i,mm^{\prime }m^{\prime \prime }m^{\prime \prime \prime
}}U_{\alpha \beta }^{mm^{\prime }m^{\prime \prime }m^{\prime \prime \prime
}}(c_{im\alpha }^{+}c_{im^{\prime \prime }\alpha }^{{}})(c_{im^{\prime
}\beta }^{+}c_{im^{\prime \prime \prime }\beta }^{{}}).
\end{equation}%
We follow Ref. \cite{AbinitioDGA} of the paper and consider the equation of motion 
\begin{eqnarray}
\lbrack G\Sigma ]_{\mathbf{k},\sigma }^{mm^{\prime }}(\tau ) &=&\langle Tc_{%
\mathbf{k}m\sigma }(\tau )[H_{\mathrm{int}},c_{\mathbf{k}m^{\prime }\sigma
}^{+}(\tau ^{\prime })]\rangle _{\tau ^{\prime }\rightarrow 0} \notag \\
&=&\sum\limits_{\alpha }\sum\limits_{i,mm^{\prime \prime }\widetilde{m}%
\widetilde{m}^{\prime }}\sum\limits_{\widetilde{\mathbf{k}}\mathbf{q}%
}U_{\alpha \sigma }^{\widetilde{m}\widetilde{m}^{\prime }m^{\prime \prime
}m^{\prime }}\langle \mathcal{T}c_{\mathbf{k}m\sigma }(\tau )c_{\widetilde{%
\mathbf{k}}\widetilde{m}\alpha }^{+}c_{\mathbf{k}+\mathbf{q},\widetilde{m}%
^{\prime }\sigma }^{+}c_{\widetilde{\mathbf{k}}\mathbf{+q,}m^{\prime \prime
}\alpha }^{{}}\rangle _{\mathrm{conn}}+\sum\limits_{\widetilde{m}}G_{%
\mathbf{k}\sigma }^{m\widetilde{m}}(\tau )\Sigma _{H}^{\widetilde{m}%
m^{\prime }}, \label{EOM}
\end{eqnarray}%
where $\mathcal{T}$ is time ordering, conn means the connected part of the two-particle Green's function, and
\begin{equation}
\Sigma _{H}^{mm^{\prime }} =\sum\limits_{mm^{\prime }\widetilde{m}%
\widetilde{m}^{\prime }}\sum\limits_{\mathbf{k}^{\prime }}\left[ -U_{\sigma
\sigma }^{m\widetilde{m},\widetilde{m}^{\prime }m^{\prime }}\langle c_{%
\mathbf{k}^{\prime }\mathbf{,}\widetilde{m}\sigma }^{+}c_{\mathbf{k}^{\prime
},\widetilde{m}^{\prime }\sigma }^{{}}\rangle +\sum\limits_{\alpha
}U_{\alpha \sigma }^{\widetilde{m}m,\widetilde{m}^{\prime }m^{\prime
}}\langle c_{\mathbf{k}^{\prime }\widetilde{m}\alpha }^{+}c_{\mathbf{k}%
^{\prime }\mathbf{,}\widetilde{m}^{\prime }\alpha }^{{}}\rangle \right] 
\label{SHF} 
\end{equation}
is the Hartree contribution to self-energy. 
From the equation (\ref{EOM}) we find the self-energy in the form%
\begin{equation*}
\Sigma _{k}^{mm^{\prime }}=\Sigma _{H}^{mm^{\prime }}+\Sigma
_{F,k}^{mm^{\prime }},
\end{equation*}%
where 
\begin{eqnarray}
\Sigma _{F,k}^{mm^{\prime }} &=&T^{2}\sum\limits_{k^{\prime
}q}\sum\limits_{\alpha }
F_{\sigma \alpha,kk^{\prime }q }^{mm^{\prime
\prime },m^{\prime \prime \prime }\widetilde{m}}
U_{\alpha \sigma }^{\widetilde{%
m}^{\prime }\widetilde{m}^{\prime \prime \prime },\widetilde{m}^{\prime
\prime }m'}
\chi _{k^{\prime }q}^{0,\widetilde{m}m^{\prime \prime },%
\widetilde{m}^{\prime }\widetilde{m}^{\prime \prime }}G_{k+q}^{m^{\prime
\prime \prime }\widetilde{m}^{\prime \prime \prime }}
\label{SigmaF}
\end{eqnarray}%
and
\begin{align}
F_{\alpha \beta,kk^{\prime }q }^{mm^{\prime },m^{\prime \prime }m^{\prime
\prime \prime }}&=G_{k,m\widetilde{m}}^{-1}G_{k^{\prime }+q,m^{\prime }%
\widetilde{m}^{\prime }}^{-1}G_{k\mathbf{+}q,m^{\prime \prime }\widetilde{m}%
^{\prime \prime }}^{-1}G_{k^{\prime }m^{\prime \prime \prime }\widetilde{m}%
^{\prime \prime \prime }}^{-1}\int d\tau _{1}d\tau _{2}d\tau _{3}e^{i\nu
\tau _{1}+i(\nu ^{\prime }+\omega )\tau _{2}-i(\nu +\omega )\tau
_{3}}\notag\\
&\times\langle \mathcal{T}c_{\mathbf{k}\widetilde{m}\sigma }(\tau _{1})c_{%
\mathbf{k}^{\prime }\mathbf{+q,}\widetilde{m}^{\prime }\alpha }^{{}}(\tau
_{2})c_{\mathbf{k}+\mathbf{q},\widetilde{m}^{\prime \prime }\sigma
}^{+}(\tau _{3})c_{\mathbf{k}^{\prime }\widetilde{m}^{\prime \prime \prime
}\alpha }^{+}(0)\rangle _{\mathrm{conn}}
\label{defF}
\end{align}%
is the interaction vertex (here and in the following, we assume summations over repeating orbital
indices);
$\chi _{k^{\prime }q}^{0,m^{\prime }m^{\prime \prime },\widetilde{m}^{\prime
}\widetilde{m}^{\prime \prime }}=-TG_{k^{\prime }}^{m^{\prime }\widetilde{m}%
^{\prime }}G_{k^{\prime }+q}^{\widetilde{m}^{\prime \prime
}m^{\prime \prime }}$  and we have introduced frequency-momenta variables $k=(i\nu_n,{\mathbf k})$, etc. The antisymmetry
of the two-particle Green's function entering Eq. (\ref{defF}) yields $F_{\alpha
\alpha,kk^{\prime }q }^{mm^{\prime
},m^{\prime \prime }m^{\prime \prime \prime }}=-F_{\alpha
\alpha,k'+q,k',k-k' }^{m^{\prime }m,m^{\prime \prime }m^{\prime \prime \prime
}}=-F_{\alpha
\alpha,k,k+q,k'-k}^{mm^{\prime },m^{\prime \prime \prime
}m^{\prime \prime }}=F_{\alpha
\alpha,k^{\prime }+q,k+q,-q}^{m^{\prime
}m,m^{\prime \prime \prime }m^{\prime \prime }}.$ We introduce 
\begin{eqnarray}
F_{c(s)} &=&-F_{\uparrow \uparrow }\mp F_{\uparrow \downarrow }   
\end{eqnarray}%
(with momenta/frequency and orbital indices omitted) and similarly for $U_{c(s)}$.
This yields%
\begin{equation}
\Sigma _{F,k}^{mm^{\prime }}=\frac{1}{2}T^{2}\sum\limits_{k^{\prime
},q}\left( F_{s,k k^{\prime } q}^{mm^{\prime \prime
},m^{\prime \prime \prime }m^{\prime }}U_{s}^{\widetilde{m}^{\prime \prime
\prime }\widetilde{m}^{\prime },\widetilde{m}\widetilde{m}^{\prime \prime
}}+F_{c,k k^{\prime } q}^{mm^{\prime \prime },m^{\prime
\prime \prime }m^{\prime }}U_{c}^{\widetilde{m}^{\prime \prime \prime }%
\widetilde{m}^{\prime },\widetilde{m}\widetilde{m}^{\prime \prime }}\right)
G_{k+q}^{m^{\prime \prime \prime }\widetilde{m}^{\prime \prime \prime }}\chi
_{k^{\prime }q}^{0,m^{\prime }m^{\prime \prime },\widetilde{m}^{\prime }%
\widetilde{m}^{\prime \prime }}.  \label{Se1}
\end{equation}%
In case of SU(2) symmetry $U_{\alpha \beta }$ can be chosen not depending on $\alpha ,\beta ,$ which yields $%
U_{s}=0,$ $U_{c}^{mm^{\prime }m^{\prime \prime }m^{\prime \prime \prime
}}=-2U^{mm^{\prime }m^{\prime \prime }m^{\prime \prime \prime }},$ and the
Eq. (\ref{Se1}) reduces to that derived in Ref. \cite{AbinitioDGA} of the paper. For this choice $U_{c(s)}$ do not correspond to the ``true" charge and spin
interaction. However, in the present paper we choose $U_{\alpha\alpha}$ antisymmetric with respect to the pairs of orbital indices (in particular, $U^{mmmm}_{\alpha\alpha}$=0), which is more convenient for Ising symmetry; the symmetric part of $U_{\alpha\alpha}$ does not give contribution to the Eq. (\ref{SigmaF}) in view of the antisymmetry of $F_{\alpha\alpha}$. In this case $U_{c(s)}$ do correspond to the interactions in
charge and spin channel. 

For the purposes of numerical evaluation, it is convenient to pick out the
local contribution from the Eq. (\ref{Se1}), which together with the
Hartree-Fock self-energy (\ref{SHF}) combines to the local self-energy $%
\Sigma _{\nu }^{\mathrm{loc,}mm^{\prime }}.$ This yields 
\begin{equation}
\Sigma _{k}^{mm^{\prime }}=\Sigma _{\nu }^{\mathrm{loc,}mm^{\prime }}+\frac{1%
}{2}T^2 \mathcal{N}_{l}\sum\limits_{k^{\prime },q}\left( F_{s,kk'q}^{mm^{\prime \prime },m^{\prime \prime \prime }m^{\prime
}}U_{s}^{\widetilde{m}^{\prime \prime \prime }\widetilde{m}^{\prime },%
\widetilde{m}\widetilde{m}^{\prime \prime }}+F_{c,kk'q}^{mm^{\prime \prime },m^{\prime \prime \prime }m^{\prime
}}U_{c}^{\widetilde{m}^{\prime \prime \prime }\widetilde{m}^{\prime },%
\widetilde{m}\widetilde{m}^{\prime \prime }}\right) G_{k+q}^{m^{\prime
\prime \prime }\widetilde{m}^{\prime \prime \prime }}\chi _{k^{\prime
}q}^{0,m^{\prime }m^{\prime \prime },\widetilde{m}^{\prime }\widetilde{m}%
^{\prime \prime }},  \label{Se1a}
\end{equation}%
where the symbol $\mathcal{N}_{l}...$ takes the non-local part of the
following expression. By performing the ``rotation" of the channels%
\begin{equation}
\widetilde{k}^{\prime }=k+q,\,\,\,\widetilde{q}=k^{\prime }-k,\,\,\,m^{\prime
}\leftrightarrow m^{\prime \prime \prime },\,\,\,\widetilde{m}^{\prime
}\leftrightarrow \widetilde{m}^{\prime \prime \prime }  \label{rot1}
\end{equation}
(such that $\widetilde{k}^{\prime }+\widetilde{q}=k^{\prime }+q$), $%
G_{k+q}^{m^{\prime \prime \prime }\widetilde{m}^{\prime \prime \prime }}\chi
_{k^{\prime }q}^{0,m^{\prime }m^{\prime \prime },\widetilde{m}^{\prime }%
\widetilde{m}^{\prime \prime }}$remains unchanged. Denoting the the
quantities in the rotated channel by tildes (e.g. $%
F_{\mu }=F_{\mu ,k k^{\prime } q}^{mm^{\prime \prime
};m^{\prime \prime \prime }m^{\prime }}$ and  $\widetilde{F}_{\mu }=F_{\mu
,k,k+q;k^{\prime }-k}^{mm^{\prime \prime };m^{\prime }m^{\prime
\prime \prime }}$) where $\mu =\alpha \beta ,c,s,$ we have according to
antisymmetry of $F_{\alpha \alpha },$ $\widetilde{F}_{\uparrow \uparrow
}=-F_{\uparrow \uparrow }.$ We also denote $\widetilde{F}_{\uparrow
\downarrow }=F_{s\perp },$ since this vertex refers to the transverse spin channel. For SU(2) symmetry of
Hund exchange the spin vertex, determined from the transverse and
longitudinal channels are identical $F_{s\perp }=F_{s},$ but for the Ising
symmetry this identity is violated. We therefore find%
\begin{equation}
\widetilde{F}_{c(s)}=\mp F_{s\perp }-\frac{F_{c}+F_{s}}{2}  \label{rot}
\end{equation}%
and similarly for $U.$ Therefore, the self-energy can be equivalently
rewritten as 
\begin{align}
\Sigma _{k}^{mm^{\prime }}=\Sigma _{\nu }^{\mathrm{loc,}mm^{\prime }}+\frac{1%
}{2}T^{2}\mathcal{N}_{l}\sum\limits_{k^{\prime },q}&\left[ (F_{s,k k^{\prime } q}^{mm^{\prime \prime },m^{\prime \prime \prime }%
\widetilde{m}}+F_{c,k k^{\prime } q}^{mm^{\prime \prime
},m^{\prime \prime \prime }\widetilde{m}})U_{\uparrow \uparrow }^{\widetilde{%
m}^{\prime \prime \prime }\widetilde{m}^{\prime },\widetilde{m}^{\prime
\prime }m^{\prime }}\right.\notag\\
&\left.+2F_{s\perp ,
k k^{\prime } q}^{mm^{\prime \prime },m^{\prime \prime \prime }\widetilde{m}}U_{s\perp }^{%
\widetilde{m}^{\prime \prime \prime }\widetilde{m}^{\prime },m^{\prime }%
\widetilde{m}^{\prime \prime }}\right] G_{k+q}^{m^{\prime \prime \prime }%
\widetilde{m}^{\prime \prime \prime }}\chi _{k^{\prime }q}^{0,\widetilde{m}%
m^{\prime \prime },\widetilde{m}^{\prime }\widetilde{m}^{\prime \prime }}.
\label{Se2}
\end{align}%
To derive the desired result for self-energy we, following to Ref. \cite{Krien} of the paper,
combine the two representations (\ref{Se1}) and (\ref{Se2}) with the weights 
$\alpha $ and $1-\alpha $ $\ $($\alpha $ is arbitrary): 
\begin{eqnarray}
&&\Sigma _{k}^{mm^{\prime }} =\Sigma _{\nu }^{\mathrm{loc,}mm^{\prime }}+%
\frac{1}{2}T^{2}\mathcal{N}_{l}\sum\limits_{k^{\prime },q}\left\{ \alpha
(F_{s,k k^{\prime } q}^{mm^{\prime \prime },m^{\prime \prime
\prime }\widetilde{m}}U_{s}^{\widetilde{m}^{\prime \prime \prime }\widetilde{%
m}^{\prime },m^{\prime }\widetilde{m}^{\prime \prime }}+F_{c,k k^{\prime } q}^{mm^{\prime \prime },m^{\prime \prime \prime }%
\widetilde{m}}U_{c}^{\widetilde{m}^{\prime \prime \prime }\widetilde{m}%
^{\prime },m^{\prime }\widetilde{m}^{\prime \prime }})\right.   \\
&&\left. +(1-\alpha )\left[ 2F_{s\perp ,k k^{\prime } q}^{mm^{\prime \prime },m^{\prime \prime \prime }\widetilde{m}}U_{s\perp }^{%
\widetilde{m}^{\prime \prime \prime }\widetilde{m}^{\prime },m^{\prime }%
\widetilde{m}^{\prime \prime }}+(F_{s,k k^{\prime } q}^{mm^{\prime \prime },m^{\prime \prime \prime }\widetilde{m}%
}+F_{c,k k^{\prime } q}^{mm^{\prime \prime },m^{\prime
\prime \prime }\widetilde{m}})U_{\uparrow \uparrow }^{\widetilde{m}^{\prime
\prime \prime }\widetilde{m}^{\prime },\widetilde{m}^{\prime \prime
}m^{\prime }}\right] \right\} G_{k+q}^{m^{\prime \prime \prime }\widetilde{m}%
^{\prime \prime \prime }}\chi _{k^{\prime }q}^{0,\widetilde{m}m^{\prime
\prime },\widetilde{m}^{\prime }\widetilde{m}^{\prime \prime }}.\notag 
\end{eqnarray}%
For Ising symmetry of Hund exchange, described by the Hamiltonian (\ref{Hint}) of the main text, the orbital indexes are pairwise equal and $U_s^{mmmm}=U^m_0$ (where $U_0$ is the intraorbital Coulomb repulsion) and $U_s^{mm'mm'}=J^{mm'}$ (where $J^{mm'}$ is Hund interaction). At the same time, the contribution of Hund exchange is absent in the transverse channel $U_{s\perp}$ because of broken SU(2) symmetry. For
correct evaluation of the transverse channel we therefore need to enforce $%
F_{s\perp }=F_{s}$ and $U_{s\perp }=U_{s}.$ Following Refs. \cite{AbinitioDGA,Krien} of the paper
we further split the vertices into the local part and the non-local ladder
contributions in the longitudinal and transverse particle-hole (ph) channels

\begin{equation*}
F_{c(s)k k^{\prime } q}^{mm^{\prime \prime },m^{\prime
\prime \prime }\widetilde{m}}=F_{c(s),\nu\nu^{\prime },\omega}^{\mathrm{loc,}mm^{\prime \prime },m^{\prime \prime
\prime }\widetilde{m}}+\mathcal{F}_{c(s),\nu\nu^{\prime },q}^{\mathrm{nl},mm^{\prime \prime },m^{\prime \prime \prime }%
\widetilde{m}}+\mathcal{F}_{c(s),\perp ,k k^{\prime } q}^{%
\mathrm{nl},mm^{\prime \prime },m^{\prime \prime \prime }\widetilde{m}}.
\end{equation*}%
(the contribution of the particle-particle channel and multi-boson terms are
assumed to be purely local). The contribution of the transverse ph channel
is further expressed as a combination of ``rotated" longitudinal channels
according to the Eq. (\ref{rot}) 
\begin{equation}
\mathcal{F}_{s(c),\perp ,k k^{\prime } q}^{\mathrm{nl}}=-%
\frac{1}{2}(\overline{\mathcal{F}}_{c,k,k+q,k^{\prime }-k}^{%
\mathrm{nl}}+\overline{\mathcal{F}}_{s,k,k+q,k^{\prime }-k}^{%
\mathrm{nl}})\mp \overline{\mathcal{F}}_{s\perp ,k,k+q,k^{\prime }-k}^{\mathrm{nl}}.
\end{equation}
where the bars mark the contributions belonging to the ``rotated" longitudinal channel.
We further use the identity%
\begin{eqnarray}
\mathcal{N}_{l}\sum\limits_{k^{\prime },q}F_{s(c),kk'q}^{mm^{\prime \prime
},m^{\prime \prime \prime }\widetilde{m}}G_{k+q}^{m^{\prime \prime \prime }%
\widetilde{m}^{\prime \prime \prime }}\chi _{k^{\prime }q}^{0,\widetilde{m}%
m^{\prime \prime },\widetilde{m}^{\prime }\widetilde{m}^{\prime \prime }} 
&=&\mathcal{N}_{l}\sum\limits_{k^{\prime },q}\left[ \mathcal{F}_{c(s),\nu
\nu ^{\prime }q}^{mm^{\prime \prime
},m^{\prime \prime \prime }\widetilde{m}}+\mathcal{F}_{c(s),\perp
,k k^{\prime } q}^{mm^{\prime \prime },m^{\prime \prime
\prime }\widetilde{m}}\right] G_{k+q}^{m^{\prime \prime \prime }\widetilde{m}%
^{\prime \prime \prime }}\chi _{k^{\prime }q}^{0,\widetilde{m}m^{\prime
\prime },\widetilde{m}^{\prime }\widetilde{m}^{\prime \prime
}}\notag\\
&&-F_{c(s),k k^{\prime } q}^{{\rm loc},mm^{\prime \prime
},m^{\prime \prime \prime }\widetilde{m}}G_{k+q}^{m^{\prime \prime \prime }%
\widetilde{m}^{\prime \prime \prime }}\left( \chi _{k^{\prime }q}^{0,%
\widetilde{m}m^{\prime \prime },\widetilde{m}^{\prime }\widetilde{m}^{\prime
\prime }}-\chi _{k^{\prime }q}^{0,\mathrm{loc,}\widetilde{m}m^{\prime \prime
},\widetilde{m}^{\prime }\widetilde{m}^{\prime \prime }}\right).   \label{Nl}
\end{eqnarray}%
In the following we neglect the contribution in the second line of Eq. (\ref{Nl}); we have
verified that it remains sufficiently small and changes quite weakly the
obtained results. Approximating again $\mathcal{F}_{s\perp %
}\simeq \mathcal{F}_{s}$ and rotating channels $\overline{%
\mathcal{F}}$ into $\mathcal{F}$ according to the Eqs. (\ref{rot1}), we
obtain%
\begin{equation}
\Sigma _{k}^{mm^{\prime }}=\Sigma _{\nu }^{\mathrm{loc,}mm^{\prime }}+\frac{1%
}{2}T^{2}\mathcal{N}_{l}\sum\limits_{k^{\prime },q}\left[ \mathcal{F}%
_{c,\nu \nu^{\prime } q}^{mm^{\prime \prime \prime},m^{\prime \prime}\widetilde{m}}U_{c}^{\widetilde{m}^{\prime \prime \prime }\widetilde{%
m}^{\prime },m^{\prime }\widetilde{m}^{\prime \prime }}+3\mathcal{F}%
_{s,\nu \nu^{\prime } q}^{mm^{\prime \prime \prime},m^{\prime \prime}\widetilde{m}}
U_{s}^{\widetilde{m}^{\prime \prime \prime }\widetilde{%
m}^{\prime },m^{\prime }\widetilde{m}^{\prime \prime }}\right]
G_{k+q}^{m^{\prime \prime }\widetilde{m}^{\prime \prime \prime }}\chi
_{k^{\prime }q}^{0,\widetilde{m}m^{\prime \prime \prime},\widetilde{m}^{\prime }%
\widetilde{m}^{\prime \prime }}
\end{equation}%
independent of $\alpha$. Finally, using 
\begin{equation}
T\sum\limits_{k^{\prime }}\mathcal{F}_{c(s),\nu\nu'q}^{mm^{\prime \prime \prime},m^{\prime \prime }\widetilde{m}}\chi
_{k^{\prime }q}^{0,\widetilde{m}m^{\prime \prime\prime },\widetilde{m}^{\prime }%
\widetilde{m}^{\prime \prime }}=\Gamma _{c(s),\nu q}^{mm^{\prime \prime },%
\widetilde{m}^{\prime }\widetilde{m}^{\prime \prime }}-1=\gamma
_{c(s),\nu q}^{mm^{\prime \prime  },\widetilde{m}%
\widetilde{m}^{\prime \prime \prime }}[1-U_{c(s)}\Pi _{c(s),q}]_
{\widetilde{m}\widetilde{m}^{\prime \prime \prime  },%
\widetilde{m}^{\prime }\widetilde{m}^{\prime \prime }}^{-1}-1
\end{equation}%
where $\Gamma_{c(s),\nu q}^{mm',m''m'''}=G^{-1}_{k,m\widetilde{m}}G^{-1}_{k+q,m'\widetilde{m}'}\langle c^{}_{k\widetilde{m}} c^{+}_{k+q,\widetilde{m}'}S^z_{q,m''m'''}\rangle_{\rm conn}$ is the triangular vertex, $\gamma_{c(s),\nu q}^{mm',m''m'''}$ is the respective reduced (Hedin) vertex, $\Pi_{c(s),q}$ is the polarization function (see, e.g., Ref. \cite{OurJq} for details), and introducing the effective interaction in the charge (spin) channel%
\begin{equation}
\lbrack 1-U_{c(s)}\Pi _{c(s),q}]_{\widetilde{m}\widetilde{m}%
^{\prime \prime \prime},\widetilde{m}^{\prime }\widetilde{m}^{\prime \prime
}}^{-1}U_{c(s)}^{m\widetilde{m}^{\prime
},m^{\prime }\widetilde{m}^{\prime \prime }}=W_{c(s),q}^{\tilde{m}m,\widetilde{m}^{\prime\prime\prime },m^{\prime }},
\end{equation}%
we find%
\begin{equation}
\Sigma _{k}^{mm^{\prime }}=\Sigma _{\nu }^{\mathrm{loc,}mm^{\prime }}+\frac{1}{2}T\mathcal{N}_{l}\sum\limits_{q}\left[ \gamma
_{c,\nu q}^{mm^{\prime \prime },\widetilde{m}\widetilde{m}^{\prime \prime
}}W_{c,q}^{\widetilde{m}\widetilde{m}^{\prime
},\widetilde{m}^{\prime \prime }m^{\prime }}+3\gamma _{s,\nu q}^{mm^{\prime
\prime },\widetilde{m}\widetilde{m}^{\prime \prime }}W_{s,q}^{\widetilde{m}\widetilde{m}^{\prime },\widetilde{m}%
^{\prime \prime }m^{\prime }}\right] G_{k+q}^{m^{\prime \prime }\widetilde{m}%
^{\prime }}. \label{EqSeS}
\end{equation}%
For pairwise equal orbital indices, considered in the density-density approximation (Eq. (\ref{Hint}) of the main text), we obtain Eq. (\ref{SEnonloc}) of the main text.

\subsection{Origin of electronic spectrum splitting}

To understand the origin of the electronic spectrum splitting in the paramagnetic case, we consider the simplified approach discussed previously for the single-orbital case \cite{QS1,QS4}. For that, we retain only the largest static ($\omega_n=0$) contributions of the spin fluctuations in the sum of Eq. (\ref{EqSeS}) and neglect momentum ${\mathbf q}$ in comparison to ${\mathbf k}$ in the argument of the electronic Green's function, assuming that the typical momenta of magnetic fluctuations $|{\mathbf q}|\ll k_F$. In this case, we represent $\Sigma _{k}^{mm^{\prime }}=\Sigma _{\mathrm{\nu }}^{\mathrm{loc},mm^{\prime}}+\tilde \Sigma^{mm'}_{k}$, where 
\begin{equation}
\tilde \Sigma _{k}^{mm^{\prime }}=
3(\Delta _{\nu }^{mm^{\prime }}G_{k}^{mm^{\prime }}-\Delta _{i\nu }^{%
\mathrm{loc},mm^{\prime }}G_{\nu }^{\mathrm{loc,}mm^{\prime }})
\end{equation}%
and 
\begin{eqnarray}
\Delta _{i\nu }^{mm^{\prime }} &=&\frac{T}{2}\gamma _{s,\nu ,q=0}^{m\widetilde{%
m}}\sum_{\mathbf{q}}W_{s,(\mathbf{q},0)}^{\widetilde{m}m^{\prime }}, \\
\Delta _{i\nu }^{\mathrm{loc,}mm^{\prime }} &=&\frac{T}{2}\gamma _{s,\nu
,\omega =0}^{{\rm loc},m\widetilde{m}}W_{s,0}^{{\rm loc},\widetilde{m}m^{\prime }}.
\nonumber
\end{eqnarray}
For $\Delta\gg \Delta^{\rm loc}$ (which implies sufficiently strong spin fluctuations) and/or large $G_k$, which is provided by the sufficiently flat electronic spectrum near the Fermi level, we obtain the resulting non-local electronic Green's function on the real frequency axis%
\begin{eqnarray}
\mathcal{G}^{\lambda \lambda'}_{k} &=&[\nu-H_{\mathbf k}+\mu+M_\mathrm{DC}-\Sigma _{k}^{mm^{\prime }}]^{-1}=
\lbrack (\nu +\mu+M_\mathrm{DC}^{\lambda }) \delta_{\lambda \lambda'}-H_{\mathbf k}^{\lambda \lambda ^{\prime }}-\Sigma _{\nu }^{\mathrm{loc},mm^{\prime
}}-\tilde \Sigma_k^{mm'}
]^{-1}, 
\end{eqnarray}
where indexes $\lambda,\lambda'$ correspond to the whole Wannier space, and $m,m'$ to the correlated $d$-orbital subspace, and
\begin{eqnarray}
\tilde \Sigma^{mm'}_k&\simeq&
3\Delta
_{\nu }^{m m^{\prime }}[(\nu +\mu+M^\lambda_\mathrm{DC})\delta_{\lambda\lambda'}-H_{\mathbf k}^{\lambda\lambda'}-\Sigma _{\nu}^{\mathrm{loc},mm^{\prime }}]^{-1}_{m m'}.
\end{eqnarray}%
Taking the $\nu\rightarrow 0$ limit of the real part of the local self-energy, supposing that $\tilde{H}_{\mathbf k}=H_{\mathbf k}+\mathrm {Re}\Sigma_{\nu\rightarrow 0}+\mu-M$ is diagonalized by the transformation $%
\tilde{H}_{\mathbf k}^{\lambda \lambda ^{\prime }}=R_{\lambda \alpha }^{+}E_{{\mathbf k},\alpha
}R_{\alpha \lambda ^{\prime }}$ and considering, for simplicity, orbital-independent $\mathrm{Im}\Sigma^\mathrm{loc}_\nu$, we find for the Green's function and the self-energy in the band indexes defined by
\begin{equation}
\mathcal{G}_{k}^{\lambda \lambda ^{\prime }}=R_{\lambda \alpha
}^{+} \mathcal{G}_{k}^{\alpha \beta}R_{\beta \lambda ^{\prime }}, \,\,\,\, \tilde{\Sigma}_{k}^{m m ^{\prime }}=R_{m \alpha
}^{+} \tilde\Sigma_{k}^{\alpha \beta} R_{\beta \lambda ^{\prime }},
\end{equation}
the Dyson equation
\begin{equation}
\mathcal{G}_{k}^{\alpha \beta}=[(\nu-E_{{\mathbf k},\alpha }-i \mathrm{Im}\Sigma^\mathrm{loc}_\nu)\delta _{\alpha \beta }-\tilde \Sigma_k^{\alpha \beta}]^{-1}_{\alpha \beta},
\end{equation}
where
\begin{equation}
\tilde{\Sigma}_{k}^{\alpha \beta}=S^{\alpha\beta}_{\gamma,\nu}(\nu-E_{{\mathbf k},\gamma}-i \mathrm{Im}\Sigma^\mathrm{loc}_\nu)^{-1},
\end{equation}
and $S^{\alpha\beta}_{\gamma,\nu}=3R_{\alpha m}R_{m\gamma }^{+}\Delta _{\nu }^{m m^{\prime
}}R_{\gamma 
m^{\prime }}R_{m'\beta}$. Near the Fermi level, one of the dispersion eigenvalues $E_{k\gamma}$ is small. Keeping only the respective largest contribution at $\nu\rightarrow 0$ to the sum over $\gamma$, we obtain 
\begin{equation}
\tilde{\Sigma}_{k}^{\alpha \beta}=S^{\alpha\beta}_{\gamma,\nu}(\nu-i \mathrm{Im}\Sigma^\mathrm{loc}_\nu)^{-1}.
\end{equation}
We note that depending on the symmetry, some of the coefficients $S^{\alpha\beta}_{\gamma,\nu}$ can be zero or small, e.g., when $\gamma$ refers to the band formed mostly by $e_g$ states, while $\alpha,\beta$ refers to the $t_{2g}$ states band in a cubic symmetric crystal. Also, irrespective of the smallness of $E_{{\mathbf k},\gamma}$, the position of the split bands 
is determined by the poles of the
Green's function,%
\begin{equation}
\det [(\nu -E_{{\mathbf k},\alpha}-i \mathrm{Im}\Sigma^\mathrm{loc}_\nu)\delta _{\alpha \beta }-S^{\alpha\beta}_{\gamma,\nu}
(\nu -E_{{\mathbf k},\gamma}-i \mathrm{Im}\Sigma^\mathrm{loc}_\nu)^{-1}
]=0.
\end{equation}%
If $\Delta _{\nu }^{\lambda \lambda ^{\prime }}\simeq \Delta _{\nu }$
depends weakly on orbital indices, $S^{\alpha\beta}_{\gamma,\nu}=3\Delta_\nu\delta_{\alpha\gamma}\delta_{\gamma\beta}$, and the poles are given by%
\begin{equation}
\nu =E_{{\mathbf k},\alpha}\pm (3\Delta _{\nu })^{1/2}+i \mathrm{Im}\Sigma^\mathrm{loc}_\nu\label{SSpl}
\end{equation}%
To understand the effect of orbital degrees of freedom, we consider a weak
deviation $\Delta _{\nu }^{\lambda \lambda ^{\prime }}=\Delta _{\nu }+\delta
\Delta _{\nu }^{\lambda \lambda ^{\prime }}$. Then we find $\nu =E_{{\mathbf k},\alpha}\pm
((3\Delta _{\nu })^{1/2}+\delta \nu _{\alpha })+i\mathrm{Im}\Sigma^\mathrm{loc}_\nu$
where 
$\delta \nu_\alpha =
(1/2)\left( {3}/{\Delta
_{\nu }}\right) ^{1/2} \delta S^{\alpha\alpha}_{\gamma,\nu}$,
and $\delta S$ is obtained by replacing $\Delta$ by $\delta \Delta$ in the definition of $S$.
Therefore, 
$\delta\nu\propto \delta \Delta/\Delta^{1/2}$, which represents a small correction to the result (\ref{SSpl}). 

In the considered approximation, we obtain the contribution to the non-local self-energy (continued to the real frequency axis) near the Fermi level $\Sigma_k \simeq \Sigma_\nu^{\rm loc}+S/(\nu-i{\rm Im}\Sigma_\nu^{\rm loc})$. This yields {\it positive} derivative $\partial {\rm Re} \Sigma_k/\partial \nu$ at small $\nu$, which is discussed in the main text. This positive derivative violates the quasi-particle concept and provides the splitting of the electronic spectrum.

\subsection{Results for CrO$_2$}

\begin{figure}[h!]
\centering
\includegraphics[width=0.5\linewidth]{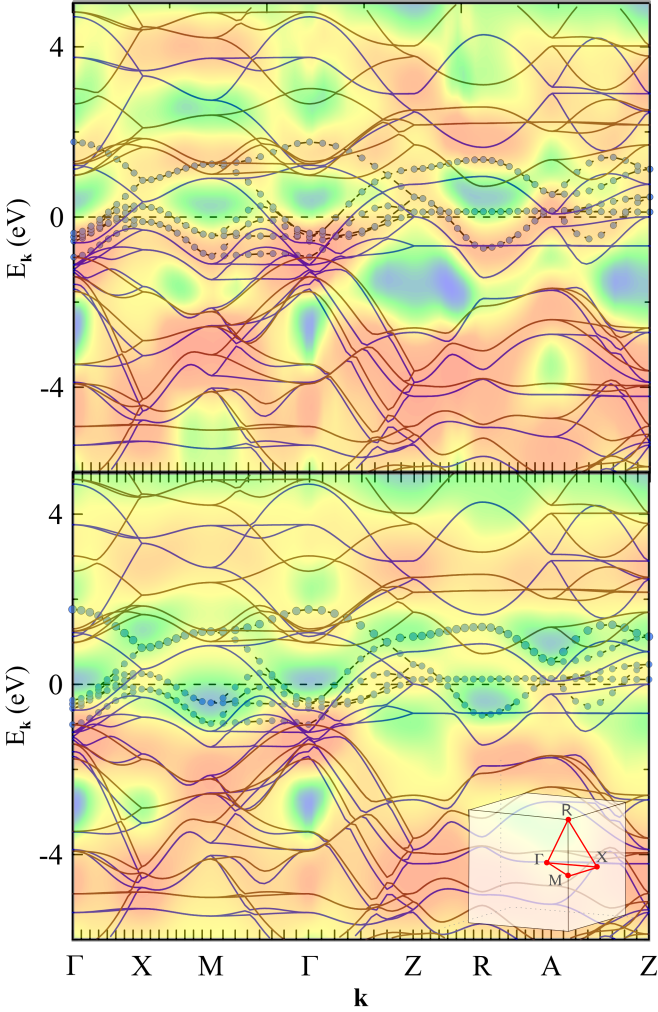}
\caption{Band structure of CrO$_2$ in the ferromagnetic phase (brown and blue solid lines for different spin projections) and low-energy (mostly $d$) states in non-magnetic phase (circles), compared to the spectral density in DFT+DMFT approach (top) and the approach, accounting the non-local corrections to the self-energy (bottom) at $\beta=6$~eV$^{-1}$.}  
\label{FigCrO2}
\end{figure}

In Fig. \ref{FigCrO2} we plot the spectral functions of DFT+DMFT and non-local approach to CrO$_2$ at $\beta=6.5$~eV$^{-1}$ in comparison with the DFT band structures for nonmagnetic and ferromagnetic phases. We use the interaction strength considered previously in Refs. \cite{MyCrO2,Solovyev_rev}, $F^0=1.99$~eV, $F^2=7.67$~eV, and $F^4=5.48$~eV. In the Wannier projection, we include both Cr $d$ and oxygen $p$ states (which results in an 11-orbital model per Cr site, see Ref. \cite{MyCrO2}), but we have verified that qualitatively similar results are obtained within the 5-orbital model, including Cr $d$ states only. The considered temperature is also chosen to be sufficiently close to the DFT+DMFT Curie temperature $\beta_c=6.8$~eV$^{-1}$ ($T_C=1700$~K). The respective uniform susceptibility at the considered temperature $\chi_{{\mathbf q}=0}=110$~eV$^{-1}$ is also much larger than the local susceptibility $\chi_{\rm loc}=12$~eV$^{-1}$,  and they are qualitatively close to those considered for iron. In the case of CrO$_2$, DMFT yields a somewhat stronger change in the electronic band structure than in iron; in particular, it shifts the flat band in the direction $Z$-$R$-$A$ away from the Fermi level, which makes the band structure similar to that obtained within DFT for the ordered phase. Considering the effect of the non-local self-energy yields a stronger splitting of bands, yielding again qualitative agreement with the ferromagnetic DFT band structure, which is related to closer proximity to half filling of the $d$ states (see main text). 

\subsection{Additional results for CrTe$_2$}

\begin{figure}[h!]
\centering
\includegraphics[width=0.60\linewidth]{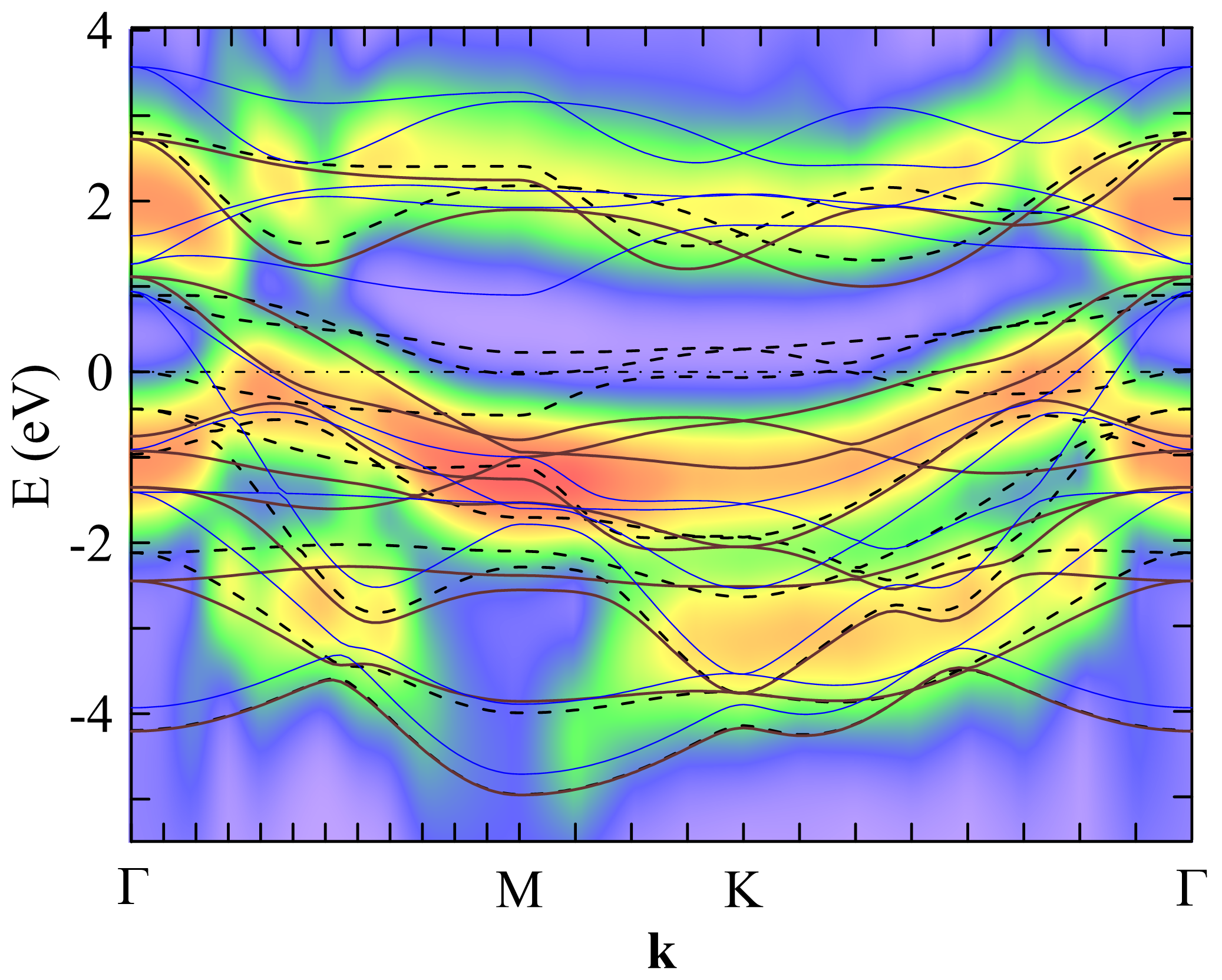}
\caption{(Color online) Band structure of monolayer CrTe$_2$ in the ferromagnetic (violet and blue solid lines for different spin projections) and non-magnetic (black dashed lines) phases, compared to the spectral density in the DFT+DMFT approach}.    
\label{Fig:CrTe2_sl_loc}
\end{figure}

\begin{figure}[h!]
\centering
\includegraphics[width=0.6\linewidth]{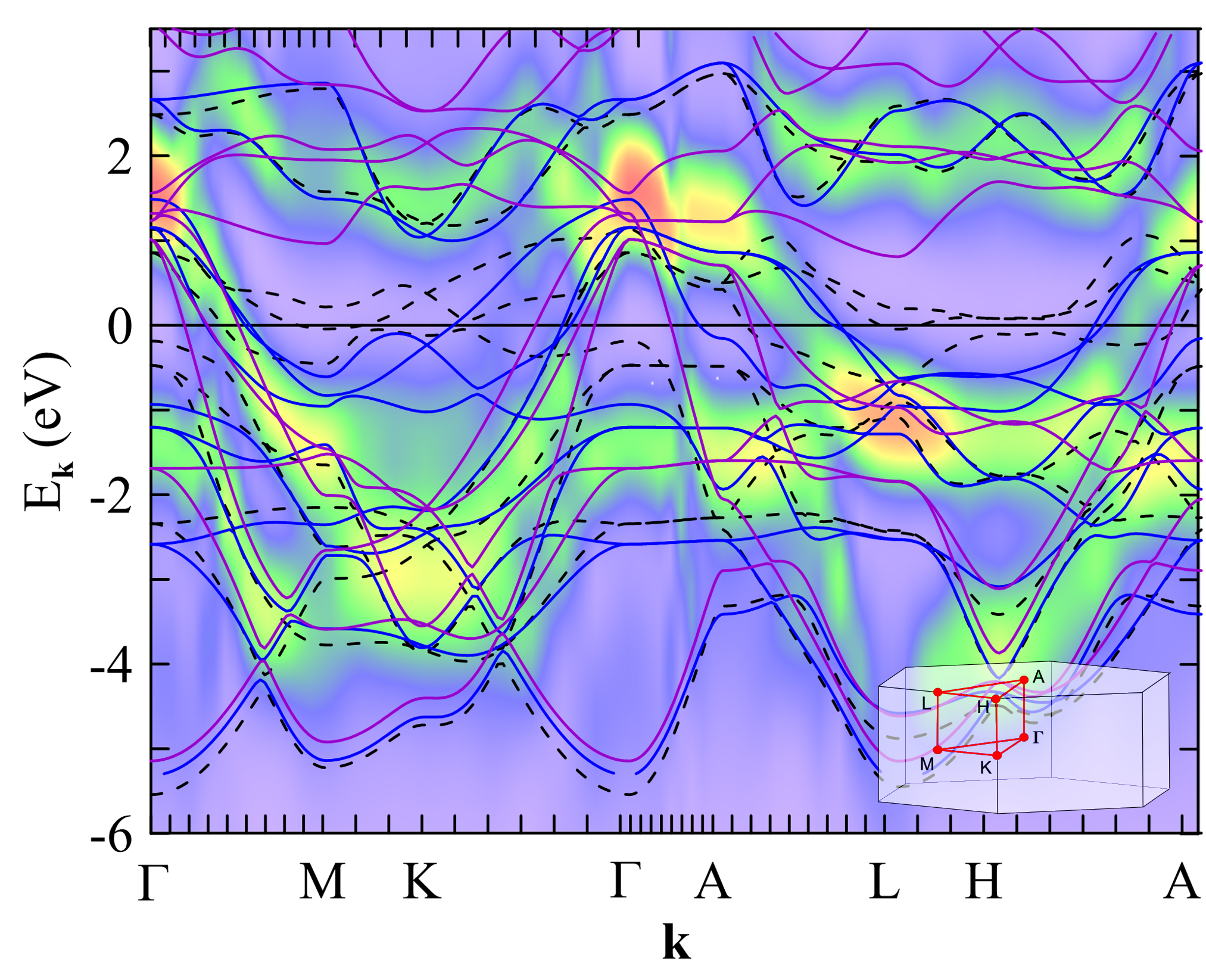}
\caption{Band structure of bulk CrTe$_2$ in the ferromagnetic (violet and blue solid lines for different spin projections) and non-magnetic (black dashed lines) phases, compared to the spectral density in DMFT approach}    
\label{Fig:CrTe2}
\end{figure}

In Fig. \ref{Fig:CrTe2_sl_loc} we present spectral functions of monolayer CrTe$_2$ at $T=1160$~K, obtained with the local self-energy. One can see that the obtained spectral functions are quite similar to those for the non-local self-energy result in Fig. \ref{Fig:CrTe2_sl} of the main text, except for the suppression of the spectral weight near the $\Gamma$ point of the Brillouin zone.

In Fig. \ref{Fig:CrTe2} we present the results for the bulk CrTe$_2$ at $T=1600$~K ($\beta=7.3$~eV$^{-1}$) which is above the magnetic transition temperature $T_c=950$~K ($\beta_c=12.2$~eV$^{-1})$; we choose the same parameters $U=F^0=2.8$~eV and $J=0.9$~eV, as for the single-layer CrTe$_2$. We obtain the maximum of the susceptibility $\chi_{{\bf  q}_A}=95$~eV$^{-1}$, which corresponds to the tendency of antiferromagnetic ordering between the layers (see Ref. \cite{CrTe2} of the main text), and $\chi_{\rm loc}=41$~eV$^{-1}$. In that case, similarly to the single-layer CrTe$_2$, considered in the main text, we find strong band splitting already in the DFT+DMFT approach, which originates from the local magnetic correlations. The effect of the non-local self-energy (not shown) is sufficiently small in this case and does not change qualitatively (and changes weakly quantitatively) the resulting band structure (we have verified that the non-local corrections remain weak also at $\beta=9$~eV$^{-1}$ closer to the DMFT phase transition). The obtained spectral functions along the $\Gamma$-$M$-$K$-$\Gamma$ direction are similar to those of the single-layer compound. 

\bibliographystyleApp{unsrt}

\end{document}